\documentclass[a4paper,12pt]{article}
\usepackage{epsfig}

\begin{document}

\def\beq{\begin{equation}}
\def\eeq{\end{equation}}
\def\bea{\begin{eqnarray}}
\def\eea{\end{eqnarray}}

\newcommand{\dedouble}{ \stackrel{ \leftrightarrow }{ \partial } }
\newcommand{\deR}{ \stackrel{ \rightarrow }{ \partial } }
\newcommand{\deL}{ \stackrel{ \leftarrow }{ \partial } }
\newcommand{\ci}{{\cal I}}
\newcommand{\ca}{{\cal A}}
\newcommand{\Wp}{W^{\prime}}

\renewcommand{\thefootnote}{\fnsymbol{footnote}}
\rightline{HIP-2003-46/TH}
\rightline{ROME1-1360/2003}
\rightline{}
\vspace{.5cm} 
{\Large
\begin{center}
{\bf Violation of Angular Momentum Selection Rules \\ 
in Quantum Gravity }

\end{center}}
\vspace{.3cm}

\begin{center}
Anindya Datta$^{1}$, Emidio Gabrielli$^{1}$, and Barbara Mele$^{2}$ \\
\vspace{.3cm}
$^1$\emph{
Helsinki Institute of Physics,
     POB 64, University of Helsinki, FIN 00014, Finland}
\\
$^2$\emph{Istituto Nazionale di Fisica Nucleare, Sezione di Roma,
and Dip. di Fisica, Universit\`a La Sapienza,
P.le A. Moro 2, I-00185 Rome, Italy}
\end{center}

\vspace{.3cm}
\hrule \vskip 0.3cm
\begin{center}
\small{\bf Abstract}\\[3mm]
\end{center}

A simple consequence of the angular momentum conservation 
in quantum field theories
is that the interference of $s$-channel amplitudes exchanging
particles with different spin $J$ vanishes after 
complete angular integration.
We show that, while this rule holds in scattering processes
mediated by a {\it massive} graviton in Quantum Gravity,
a {\it massless} graviton $s$-channel exchange
breaks orthogonality when considering its interference with a scalar-particle 
$s$-channel exchange, whenever all the external states are  massive.
As a consequence, we find that, in the Einstein theory, unitarity
implies that angular momentum is not conserved 
at quantum level in the graviton 
coupling to massive matter fields.
This result can be interpreted as a new anomaly, revealing unknown aspects of  
the well-known van Dam - Veltman - Zakharov discontinuity.

\begin{minipage}[h]{14.0cm}
\end{minipage}
\vskip 0.3cm \hrule \vskip 0.5cm
\section{Introduction}

It is well known that, when considering a massive spin-2 gravitational 
field in quantum gravity, 
the limit of vanishing graviton mass  
is  distinct from the prediction of the massless-graviton
Einstein theory.
In \cite{vdv}, \cite{zakh}, van Dam, Veltman,  and Zakharov (vDVZ)
stressed this problem considering the leading tree-level approximation
to the graviton exchange between matter sources,
for a massive graviton coupled to matter as $h^{\mu \nu}
T_{\mu \nu}$  (with $T_{\mu \nu}$ the conserved energy-momentum tensor
and $h^{\mu \nu}$ the graviton field).
The vDVZ discontinuity is shown to arise from the fact that
a massive spin-2 tensor field has  five 
polarization degrees of freedom, while a massless spin-2 graviton
has simply two.
In the massless limit, the massive graviton decomposes into three 
massless fields with spin-2, 
spin-1 and spin-0, respectively. The spin-1 vector field has a 
derivative coupling to the conserved energy-momentum tensor,
and its contribution to the one graviton exchange amplitude vanishes.
On the other hand, the spin-0 scalar field is coupled to the trace of the 
energy-momentum tensor and contributes in general to the 
scattering amplitude. 
This scalar component does not decouple even in the massless graviton limit.
This gives rise to a discontinuity in the predictions of the massive 
and massless theory in the lowest tree-level approximation.
As a consequence, in the massive theory (even in 
the limit of small masses)
the light bending by the Sun and the precession of the Mercury perihelion
differ by numerical factors from the predictions of the Einstein theory.

Many papers have elaborated on the possibility to fix this apparent 
inconsistency of the massive theory, in different directions
\cite{van}-\cite{ddgv}.
For instance, in \cite{van} it is claimed that, 
if the light bending by the sun
is computed by solving the exact space-time metric equation in 
the presence of a small graviton mass, no discontinuity 
arises in the limit of small graviton mass. In fact,
the discontinuity could be connected to the use 
of perturbation theory for the metric fluctuations around the flat space-time.
More recently, it has been shown that there is not 
any vDVZ discontinuity in the De Sitter space \cite{higuchi} (or in the 
Anti De 
Sitter space \cite{ads}), where the massless graviton limit is smooth
(see also  \cite{dgm}, \cite{ddgv} for other  solutions).

Here, we present a different class of problems connected
to the vDVZ discontinuity. 
In particular, we stress the fact that there are cases where, 
while the massive theory is well-behaved,
a massless graviton gives rise to  inconsistencies.
In particular, we show that the massless graviton 
propagator in the Einstein theory breaks angular momentum selection rules. 
\begin{figure}[t]
\begin{center} 
\hspace*{0mm}
\epsfig{file=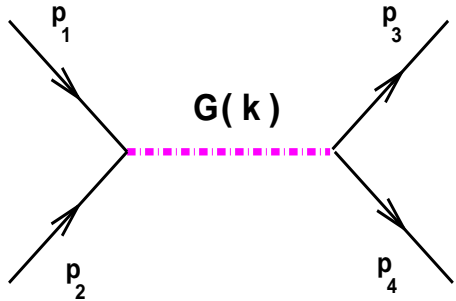,width=6cm,height=4cm}\\
\caption{{\small Scattering $p_1\, p_2\to p_3\, p_4$ in the $s$-channel
with a graviton exchange.
 }}
\label{fig1}
\end{center}
\end{figure}

Let us consider the tree-level amplitude for the graviton
exchange in the $s$-channel between two on-shell matter fields 
(Fig. \ref{fig1}).
The two on-shell matter fields enter into the amplitude
through the conserved (at the zeroth order in $h_{\mu\nu}$) symmetric
energy-momentum tensors $T_{\mu \nu}$ and $T^{\prime}_{\alpha \beta}$,
respectively\footnote{
In this paper, indices ($\mu,\nu,\alpha,\beta$) are contracted according 
to the Minkowski metric 
$\eta_{\mu\nu}={\rm Diag}(1,-1,-1,-1)$.}.

For a {\it massive}  spin-2 field of momentum $k$ and mass $m_G$, 
one has five independent
polarization tensors $\epsilon_{\mu\nu}(k,\sigma)$, where
the index $\sigma$ runs over the polarization states.
Summing over all polarizations, one gets \cite{vdv}
\beq
\sum_{\sigma=1}^{5} \epsilon_{\mu\nu}(k,\sigma)\, \epsilon_{\alpha\beta}
(k,\sigma)=
P^m_{\mu\nu\alpha\beta}(k)
\eeq
with
\bea
P^m_{\mu\nu\alpha\beta}(k)&=&\frac{1}{2}\left(\eta_{\mu\alpha}\eta_{\nu\beta}
+\eta_{\mu\beta}\eta_{\nu\alpha} -\eta_{\mu\nu}\eta_{\alpha\beta}\right)
\nonumber\\
&-&\frac{1}{2\, m_G^2}\left(\eta_{\mu\alpha}k_{\mu}k_{\beta}+
\eta_{\nu\beta}k_{\mu}k_{\alpha}+
\eta_{\mu\beta}k_{\nu}k_{\beta}+
\eta_{\nu\alpha}k_{\mu}k_{\beta}\right)
\nonumber\\
&+&\frac{1}{6}\left(\eta_{\mu\nu}+\frac{2}{m_G^2}k_{\mu}k_{\nu}\right)
\left(\eta_{\alpha\beta}+\frac{2}{m_G^2}k_{\alpha}k_{\beta}\right) .
\label{pol}
\eea
The projector $P^m_{\mu\nu\alpha\beta}$ is 
symmetric and traceless in both $(\mu, \nu)$ and $(\alpha, \beta)$ indices,
and
satisfies the transversality conditions
$k^{\mu} P^m_{\mu\nu\alpha\beta} = k^{\alpha} P^m_{\mu\nu\alpha\beta}=0$.

For a massless graviton, one has just  two transverse 
polarization states $(\sigma=1,2)$, that 
correspond  to the helicity values $\lambda=\pm 2$.
The sum over polarizations is then \cite{vdv}
\beq
\sum_{\sigma=1}^{2} \epsilon_{\mu\nu}(k,\sigma)\, \epsilon_{\alpha\beta}
(k,\sigma)=
P_{\mu\nu\alpha\beta}(k)=
\frac{1}{2}\left(
\eta_{\mu\alpha}\eta_{\nu\beta}+\eta_{\mu\beta}\eta_{\nu\alpha}-
\eta_{\mu\nu}\eta_{\alpha\beta}
\right) + \dots \; ,
\label{pol_G}
\eeq
where  dots stand for terms containing at least one 
graviton momentum.

In the unitary gauge, the corresponding massive and
massless graviton propagators are proportional to the projectors  
$P^m_{\mu\nu\alpha\beta}$ and  $P_{\mu\nu\alpha\beta}$, respectively
\cite{vdv}.
However, terms proportional to the graviton momentum in Eqs.(\ref{pol}) and 
(\ref{pol_G}) vanish when contracted with 
$T_{\mu\nu}$ in the on-shell matrix elements, due to the conservation of 
the energy-momentum tensor.
For this reason, the tree-level diagram with one graviton exchange
in Fig.\ref{fig1} is gauge invariant, and the effective
massive and massless graviton propagators become \cite{vdv}
\bea
G^m_{\mu\nu\alpha\beta}(k)&=&i\, \;
\frac{
\frac{1}{2} \eta_{\mu\alpha}\eta_{\nu\beta}
+ \frac{1}{2} \eta_{\mu\beta}\eta_{\nu\alpha} 
- \frac{1}{3} \eta_{\mu\nu}\eta_{\alpha\beta}}
{k^2-m^2_G+i\epsilon}
\label{pro}\\
G_{\mu\nu\alpha\beta}(k)&=&i\, \;
\frac{
\frac{1}{2} \eta_{\mu\alpha}\eta_{\nu\beta}
+ \frac{1}{2} \eta_{\mu\beta}\eta_{\nu\alpha} 
- \frac{1}{2} \eta_{\mu\nu}\eta_{\alpha\beta}}
{k^2+i\epsilon}
\label{pro_G}.
\eea
As shown in \cite{vdv}, unitarity fixes uniquely the coefficients of the
Minkowski metric products in Eqs.(\ref{pro}) and (\ref{pro_G}).

The corresponding on-shell $s$-channel matrix elements will be then, 
up to some coupling constant,
\beq
{\cal A}^m \sim T^{\mu\nu} \; G^m_{\mu\nu\alpha\beta}(k) \; 
T^{\prime\, \alpha\beta} 
\label{amp}
\eeq
and 
\beq
{\cal A} \sim T^{\mu\nu} \; G_{\mu\nu\alpha\beta}(k) \; 
T^{\prime\, \alpha\beta} 
\label{amp_G}
\eeq
In the limit $m_G\to 0$, Eqs. (\ref{amp}) and (\ref{amp_G})
only differ by the coefficients of the
$\eta_{\mu\nu}\eta_{\alpha\beta}$ term in Eqs.(\ref{pro}) and (\ref{pro_G}).
When contracted with the energy-momentum tensors, the latter
give terms proportional to the traces $T_{\mu}^{\mu}$ and
$T^{\prime\alpha}_{\alpha}$, that are nonvanishing for massive external
fields.
From this difference, the vDVZ discontinuity arises \cite{vdv}. 

Note that the  terms in the amplitudes corresponding
to the $\eta_{\mu\nu}\eta_{\alpha\beta}$ terms in the graviton propagators 
can be interpreted as a scalar field exchange amplitude
\footnote {The different coefficients of the $\eta_{\mu\nu}\eta_{\alpha\beta}$
 term in the massive and
massless graviton propagators is usually interpreted as an extra spin-0 field,
corresponding to one of the five polarization states of a massive graviton 
contributing to the massive-graviton  amplitude in the limit
$m_G\to 0$, as discussed above.}.

Let us consider now the interference of the  $s$-channel amplitudes
exchanging particles of  different spin $J$ (Fig. \ref{fig2}).
\beq
{\cal I}(i,j) \sim \; {\cal A}^{\star}(J=i)\;\times \; {\cal A}(J=j)  
\, + h.c.\, \; \; \; \; \; \;  
(j \neq i)\, 
\eeq
A simple consequence of angular momentum conservation is that,
after complete angular integration on the  final state, this quantity
must vanish, that is
\beq
\int d\cos\theta \; d\varphi \; \; {\cal I}(i,j) = 0  
\; \; \; \; \; \;  (j \neq i)\, , 
\label{orth}
\eeq
where $\theta$ is the scattering angle and $\varphi$ is the azimuthal angle
in the center of mass frame.
For instance, 
it is straightforward to verify this in gauge theories, looking at
the interference of a vector boson exchange with a scalar (Higgs boson)
particle exchange.
 
One then expects the same is true  for the interference of the 
$J=2$ and $J=0$ amplitudes.
On the other hand, we have seen above that (in the small
$m_G$ limit) the massive and massless
graviton propagator effectively differs by a scalar field exchange,
when the external fields are massive.
This extra scalar field component, when interfering with a spin-0 exchange
amplitude, will give a nonvanishing contribution to 
$\int d\cos\theta \; d\varphi \; {\cal I}(2,0)\;$.
This implies that the 
orthogonality condition in Eq. (\ref{orth})
for the interference ${\cal I}(2,0)$ can be verified 
{\it either} for the massive graviton exchange {\it or}
for the massless graviton exchange, but can NOT hold in {\it both}
cases at the same time.
  
We checked the above statement by an explicit calculation.
The result is that the orthogonality condition in Eq.(\ref{orth})
holds for the {\it massive} graviton exchange, but not 
in the Einstein theory !
\\ For a massless graviton and massive external states, one finds
\beq
\int d\cos\theta \; d\varphi \; \; {\cal I}(2,0) \; \neq \; 0   .
\label{north}
\eeq

In the following, we illustrate  this result, by giving the explicit
expressions of the above discontinuity for the scattering of different external
states. We will also extend the discussion to the 
interferences of the graviton graphs with vector-boson exchange 
diagrams in the $s$ channel.
As a theoretical 
framework, we assume the Standard Model minimally coupled
to gravity (e.g., as in \cite{QG}).

\section{ The Graviton-Scalar Interference}
\begin{figure}[t]
\begin{center} 
\hspace*{0mm}
\epsfig{file=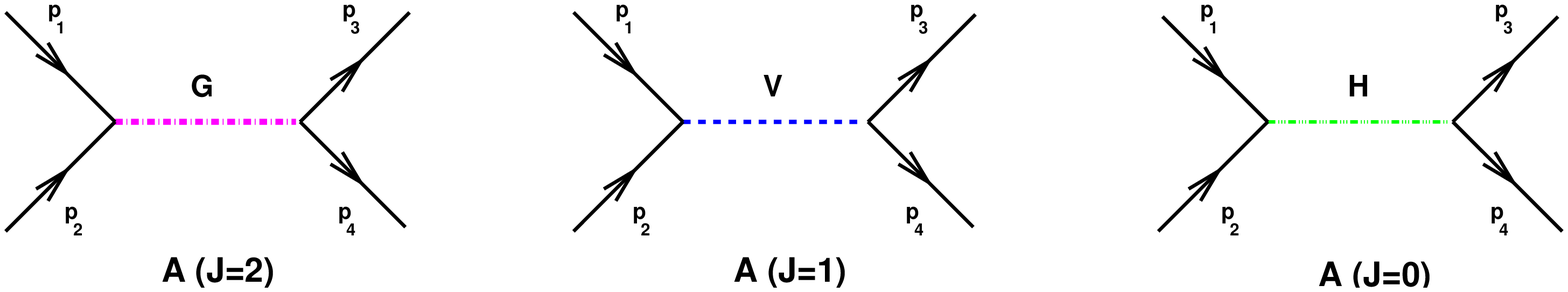,width=13cm,height=4cm}\\
\caption{{\small 
Scattering  $p_1\, p_2\to p_3\, p_4$ in the $s$-channel 
with different spin-$J$ particle exchange.}}
\label{fig2}
\end{center}
\end{figure}

In the following, we will discuss the interference of 
the on-shell tree-level scattering amplitudes in the $s$-channel 
mediated by a graviton ($J=2$) with 
either  a scalar particle exchange ($J=0$) or a vector particle
exchange ($J=1$), as in Fig. \ref{fig2}. 
We consider initial and final states containing either massive
fermions or massive vector bosons.
For each $s$ scattering channel,
\beq
a + \bar a \to  b + \bar b ,
\eeq
it is convenient to introduce
the dimensionless quantities $\;\ci^m_{a,b}(2,j)\;$ and $\;\ci_{a,b}(2,j)\;$ 
connected to the interferences of the massive and massless 
graviton amplitudes, ${\cal A}^m _{a,b}(J=2)$ and ${\cal A}_{a,b} (J=2)$, 
respectively, and the amplitude mediated by
a particle of spin $j\;$, $\;\ca_{a,b}(J=j)$, with $j=0,1$.
\\
 The  crucial point is that
the two  amplitudes ${\cal A}^m_{a,b} (J=2)$ and  
${\cal A}_{a,b} (J=2)$
 depend on the two different (massive or massless) graviton propagators in 
Eqs.(\ref{pro}) and (\ref{pro_G}),
respectively.
\\
By setting $r_{j}=m_{j}^2/s$ [with $m_j=m_{0} \; (m_{1})$ 
for the exchange of a 
scalar (vector) particle of mass $m_{0} \; (m_{1})$]
and $r_{G}=m_{G}^2/s$, with $\sqrt{s}$ 
the c.m. scattering energy, we define
\bea
\ci^m_{a,b}(2,j)
&\equiv& \frac{M_P^2}{s}(1-r_j)(1-r_G)
\sum_{\rm pol} \ca_{a,b}^{\star}(J=j)\times \ca_{a,b}^m (J=2) + h.c. \; ,
\label{iim} \\
\ci_{a,b}(2,j)
&\equiv& \frac{M_P^2}{s}(1-r_j)
\sum_{\rm pol} \ca_{a,b}^{\star}(J=j)\times \ca_{a,b} (J=2) + h.c. \; ,
\label{ii}
\eea
where $M_P$ is the reduced Planck mass (see Appendix I), 
and a sum over all the external particles polarization states is performed.
\\
Note that, by definition, 
the quantities $\ci^m_{a,b}(2,j)$ and $\ci_{a,b}(2,j)$
depend neither on the masses of particles exchanged
in the propagators nor on the Plank mass.
\\
Since we are interested into the discontinuity in the massive and massless
graviton interferences, 
 it is useful to define also the
quantity $\Delta_{a,b}(2,j)$, 
\beq
\Delta_{a,b}(2,j)\; \equiv \;\; \ci_{a,b}(2,j) - \ci^m_{a,b}(2,j) \; ,
\label{disc}
\eeq 
that gives the {\it excess} in the Einstein interference $\ci_{a,b}(2,j)$
with respect to the massive graviton interference $\ci^m_{a,b}(2,j)$
[when $j=0$, $\Delta_{a,b}(2,j)$ will be directly connected to the 
vDVZ discontinuity].

Following the  discussion in the previous section, 
we now concentrate on the graviton interference with a scalar particle,
and express all our results in terms of 
the massive graviton interference
$\ci^m_{a,b}(2,0)$ and the discontinuity $\Delta_{a,b}(2,0)\;$. 
In the $J=0$ propagator,
we assume as a  scalar particle 
 a Higgs  boson,  coupled as in the standard model (see Appendix I ). 
The following external states are  considered\footnote{
We consider only processes that do not receive contributions from 
$t~(u)$ channel exchanges.} : \\
a) the scattering of two electrons into a pair of fermions $f$, with $f\neq e$;
\\
b) the scattering of two electrons into a pair of gauge vector bosons $W$ ;
\\
c) the scattering of two  $W$'s into a pair of 
gauge vector bosons $W^{\prime}$, with $W^{\prime}\neq W$.

In the following,  $r_{i}=m_{i}^2/s$,
$\;\beta_{i}=\sqrt{1-4 r_{i}}\;\;$ ($i=e,f,W,W^{\prime}$), and
$\lambda_{e}$ $(\lambda_{f})$ is the $e$ $(f)$ Yukawa coupling. 
The angle $\theta$ is the scattering angle of a final
particle with given electric  charge with respect to the initial 
particle of same charge, in the c.m. system.
\\
Following the Feynman rules in Appendix I,
one then gets\footnote{
Results in Eqs.(\ref{feru}) and (\ref{fWa}) were first obtained 
in \cite{dgm2}, although in a different context.}
\begin{itemize}
\item
${\bf e^+ e^- \to f\bar{f}}$
\bea
\ci^m_{e,f}(2,0)&=& -\frac{8}{3} \lambda_e\lambda_f 
\beta_e^2\beta_f^2 \, \sqrt{r_e r_f}\left(1-3\cos^2{\theta}\right)
\label{feru}
\eea
and 
\beq
\Delta_{e,f}(2,0)= -\frac{4}{3}\lambda_e\lambda_f\beta_e^2\beta_f^2
\, \sqrt{r_e r_f}
\label{ferd}
\eeq
\item
${\bf e^+ e^- \to W^+ W^- }$
\bea
\ci^m_{e,W}(2,0)&=& -\frac{1}{3}\lambda_e g_{W}\sqrt{\frac{r_e}{r_W}}
\beta_e^2\beta_W^2 \, \left(1+6 r_W\right)\left(1-3\cos^2{\theta}\right)
\label{fWa}
\eea
and
\beq
\Delta_{e,W}(2,0)= \frac{1}{12}
\lambda_e g_{W}\sqrt{\frac{r_e}{r_W}}
\beta_e^2\left(3+\beta_W^2\left(1-12 r_W\right)\right)
\label{fWd}
\eeq
\item
${\bf W^+ W^- \to W^{\prime +} W^{\prime -} }$
\bea
\ci^m_{W,W^{\prime}}(2,0)
&=& -\frac{1}{24}\frac{g_{W}g_{W^{\prime}}}{\sqrt{r_W r_{W^{\prime}}}}
\beta_W^2\beta_{W^{\prime}}^2
\left(1+6 r_W\right)\left(1+6r_{W^{\prime}}\right)\,
\left(1-3\cos^2{\theta}\right) \;\;
\label{WWa}
\eea
and 
\beq
\Delta_{W,W^{\prime}}(2,0)= -\frac{1}{12}
\frac{g_{W}g_{W^{\prime}}}{\sqrt{r_W r_{W^{\prime}}}}
\beta_W^2\beta_{W^{\prime}}^2
\left( \beta_W^2 + 12 r_W^2 \right)\,
\left( \beta_{W^{\prime}}^2 + 12 r_{W^{\prime}}^2 \right) .
\label{WWd}
\eeq
\end{itemize}

Then, in each of the above channels, we have for the 
graviton-scalar interference in
the Einstein theory
\beq
\ci_{a,b}(2,0) = \ci^m_{a,b}(2,0) + \Delta_{a,b}(2,0) ,
\label{res}
\eeq
with a $\theta$ {\it independent} discontinuity $\Delta_{a,b}(2,0)$.
 
The angular integration $\int d\cos\theta$ of all the {\it
massive}
graviton interferences, $\ci^m_{a,b}(2,0)$, has  a vanishing results 
(respecting angular momentum selection rules).
On the other hand, the angular integration of 
the {\it massless} graviton interference always  gives rise to a 
nonnull results (for massive external states), that is
\beq
\int^1_{-1} d\cos\theta \; \ci_{a,b}(2,0) = \int^1_{-1} d\cos\theta \; 
\Delta_{a,b}(2,0) 
= \; 2 \; \; \Delta_{a,b}(2,0) \neq 0 \;,
\label{resdue}
\eeq
that is  connected to the vDVZ discontinuity.
 
Note that the results above do not depend on the gauge choice. For instance,
in a covariant gauge, the gauge dependence  affects 
 the  graviton propagators only through momentum
dependent terms, 
that vanish after contraction with the
energy-momentum tensors.

In Eqs.(\ref{feru})-(\ref{fWd}), the interferences are all vanishing
in the massless fermion limit ($r_{e,f} \to 0$),
due to fermion chirality.
The $J=2$ amplitude conserves the chirality, 
while the opposite is true for the $J=0$ scalar channel.
Then, in order to get a nonvanishing result for the interference, 
a chirality flip is needed in the initial/
final fermion states,  giving rise to the fermion mass factor.
In Eqs.(\ref{fWa})-(\ref{WWd}), the singularity in the external 
gauge-boson mass ($1/\sqrt{r_W}$ and $1/\sqrt{r_W^{\prime}}$ terms) 
arises from the sum over the gauge bosons polarizations, 
since longitudinal modes do not decouple in the 
massless gauge boson limit \footnote{Note that 
the $s$-channel diagram mediated by a scalar particle
with external gauge bosons does not exist in
the gauge symmetric phase, but only after spontaneous symmetry breaking.}.
 
From the  results above, assuming angular momentum  conservation
at each interaction vertex, 
one could conclude that 
the Einstein graviton propagator behaves as if it
was propagating a further scalar degree of freedom that is coupled 
to the masses of external states.
However, this would be in contrast with unitarity and  the conservation
of the energy momentum tensor. Indeed, only 
the spin-2 transverse polarizations $\epsilon_{\mu\nu}(k,\sigma)$
with helicities $\lambda =\pm 2$ are effectively exchanged 
in the massless graviton propagator (see \cite{vdv} for details).
Then, in the Einstein theory, unitarity
implies that angular momentum is not conserved 
at quantum 
level in the graviton coupling 
to massive matter fields, even if the total angular momentum is conserved
in the scattering process.

\vskip 0.5cm

We checked the results relative to the fermion-fermion scattering
by computing the  expansion in terms 
of spherical harmonics (i.e., the angular momentum eigenstates, 
${\bf Y_{l}^{m}}(\theta,\varphi)$,  defined in the Appendix I )
of the scattering amplitudes, for the four-fermion processes 
\beq
e^+(p_1,\nu_1)\,+ \, e^-(p_2,\nu_2)\, \to \, (J=0,1,2) \, \to\,
\bar{f}(p_3,\nu_3)\,+ \, f(p_4,\nu_4) 
\label{fer}
\eeq
where 
a virtual particle of spin $J=0,1,2$ is exchanged in the $s$ channel, and
$p_i$ and $\nu_i$ ($i=1,2,3,4$) stands for the external particles  momenta and
helicities, respectively.
We will work in 
the c.m. frame, where the momenta $p_i$ can be cast in the following form 
\bea
p_1&=&\frac{\sqrt{s}}{2}(1,0,0,\beta_e),~~~
p_2=\frac{\sqrt{s}}{2}(1,0,0,-\beta_e),
\nonumber \\
p_3&=&\frac{\sqrt{s}}{2}(1,\beta_f\sin{\theta}\cos{\varphi},
\beta_f\sin{\theta}\sin{\varphi},\beta_f\cos{\theta}),
\nonumber \\
p_4&=&\frac{\sqrt{s}}{2}(1,-\beta_f\sin{\theta}\cos{\varphi},
-\beta_f\sin{\theta}\sin{\varphi},-\beta_f\cos{\theta})\, ,
\eea
with $\varphi$ being the azimuthal angle.

In order to express the $J=0, 1, 2$ helicity amplitudes
as a linear combination of the spherical harmonics 
${\bf Y_{l}^{m}}(\theta,\varphi)$, 
it is convenient to use
the solution of the Dirac equation
for the particle ($U$) and antiparticle ($V$) bispinors in 
the momentum space \cite{landau}
\bea
U(p,\nu)=\left ( \begin{array}{c} 
\sqrt{\epsilon+m}\,\,  \omega_{\nu}(\underline{n})
\\ 
\sqrt{\epsilon-m}\,\,  
\left(\underline{\sigma}\cdot \underline{n}\right) \omega_{\nu}(\underline{n})
\end{array} 
\right)
~~~~~
V(p,-\nu)=\left ( \begin{array}{c} 
\sqrt{\epsilon-m}\,\, 
\left(\underline{\sigma}\cdot \underline{n}\right) \omega_{\nu}(\underline{n})
\\ 
\sqrt{\epsilon+m}\,\,  \omega_{\nu}(\underline{n}) 
\end{array} 
\right)  ,
\label{spi}
\eea
where the 2-component spinors $\omega_{\nu}(\underline{n})$ 
(with $\nu=\pm 1$) are the eigenstates of the helicity operator
$\left(\underline{\sigma}\cdot \underline{n}\right) 
\omega_{\nu}(\underline{n})
=\nu\, \omega_{\nu}(\underline{n})$, and
$\; \sigma_i$ are the Pauli matrices. 
Here, $\underline{n}=\underline{p}/|\underline{p}|$, where
$\underline{p}$ is the 3-momentum
$\underline{p}=|\underline{p}|\left(\sin{\theta}\cos{\varphi},
\sin{\theta}\sin{\varphi},\cos{\theta}\right)$, and
$\epsilon$ is the corresponding energy.
In polar coordinates, $\omega_{\nu}(\underline{n})$ can be expressed as
\bea
\omega_{+1}(\underline{n})=\left ( \begin{array}{c} 
e^{-i\frac{\varphi}{2}}\, \cos{\frac{\theta}{2}}
\\ 
e^{i\frac{\varphi}{2}}\, \sin{\frac{\theta}{2}}
\end{array} 
\right)\, ,
~~~~~
\omega_{-1}(\underline{n})=\left ( \begin{array}{c} 
-e^{-i\frac{\varphi}{2}}\, \sin{\frac{\theta}{2}}
\\ 
e^{i\frac{\varphi}{2}}\, \cos{\frac{\theta}{2}}
\end{array} 
\right)\, .
\eea
After some straightforward algebra, the ${\cal A}(J=0,1,2)$ helicity
amplitudes for the channels in Eq.(\ref{fer})
can be cast in the following form,
as a function of the initial and final helicities ($\nu_i=\pm1$)
\footnote{We do not include the axial coupling in the 
${\cal A}(J=1)$ amplitude, since the latter does not affect 
the discontinuity.},
\bea
{\cal A}(J=0)&=&R_0\, \left\{\, \delta_{\nu_1,\nu_2}
\delta_{\nu_3,\nu_4} \nu_1\nu_3 \, {\bf Y_0^0}\, 
\right\} \; ,
\label{ampz}
\\
\nonumber \\
{\cal A}(J=1)&=&R_1\, \left\{
\, \delta_{\nu_1,-\nu_2} \delta_{\nu_3,-\nu_4}
\left({\bf Y_1^0}+\sqrt{3}\, \nu_1\nu_3\,  {\bf Y_0^0}
\right)\right.
\nonumber\\
&-&\left.
\delta_{\nu_1,-\nu_2} \delta_{\nu_3,\nu_4}
\left(2\sqrt{2r_f}\, \nu_1\nu_3\,\, {\bf Y_{1}^{\bf \nu_{\bf 1}}}\right) 
-\delta_{\nu_1,\nu_2} \delta_{\nu_3,-\nu_4}
\left(2\sqrt{2r_e}\,\, {\bf Y_{1}^{-{\bf \nu_{\bf 1}}}}\right) 
\right.
\nonumber\\
&+&
\left.
\delta_{\nu_1,\nu_2} \delta_{\nu_3,\nu_4}
\left(4\nu_1\nu_3\sqrt{r_er_f}\,\, {\bf Y_1^0}\right)\, 
\right\} \; ,
\label{ampu}
\\
\nonumber
\\
{\cal A}^{\xi}(J=2)&=&R_2 \, \left\{\, 
\delta_{\nu_1,-\nu_2} \delta_{\nu_3,-\nu_4}
\left(4 \, {\bf Y_2^0}+\sqrt{5}\left({\bf Y_0^0}-\sqrt{3}\, 
\nu_1\nu_3\, {\bf Y_1^0}\right)\right)
\right.
\nonumber\\
&+&\left.
\delta_{\nu_1,-\nu_2} \delta_{\nu_3,\nu_4}
\left(\,4\sqrt{6r_f}\,\ \nu_1\nu_3\, {\bf Y_{2}^{\bf \nu_{\bf 1}}}
\right) 
+\delta_{\nu_1,\nu_2} \delta_{\nu_3,-\nu_4}
\left(\,4\sqrt{6r_e}\,\,{\bf Y_{2}^{-{\bf \nu_{\bf 1}}}}\, 
\right) 
\right.
\nonumber\\
&+&
\left.
\delta_{\nu_1,\nu_2} \delta_{\nu_3,\nu_4}
\left(8 \sqrt{r_er_f}\,
\nu_1\nu_3\left(2 \, {\bf Y_2^0}-\sqrt{5}\left(1-3\xi\right)\, 
{\bf Y_0^0}\right)\right)\, 
\right\}\; ,
\label{amplitudes}
\eea
where $\delta_{\nu_i,\nu_j}=1$  if $\nu_i=\nu_j$ and zero otherwise,
\bea
R_0=\sqrt{4\pi}\, \frac{\lambda_e\lambda_f}{1-r_0}
\beta_e\beta_f,~~~~
R_1=\sqrt{\frac{\pi}{12}}\, \frac{g_V^eg_V^f}{1-r_1},~~~~
R_2=
-\frac{1}{12}\sqrt{\frac{\pi}{5}}\, \left(\frac{s}{M_P^2}\right)
\frac{\beta_e\beta_f}{1-r_G}\, .
\nonumber
\eea
In the  ${\cal A}^{\xi}(J=2)$ graviton amplitude, the quantity
$\xi$ 
parametrizes the vDVZ discontinuity,
with  $\xi=\frac{1}{3}$ and $\xi=\frac{1}{2}$
for the massive and massless graviton propagator, respectively.
The functions ${\bf Y_l^m}(\theta,\varphi) \;$ (note that
the relevant ones are reported in Appendix I)
satisfy the following normalization condition
\beq
\int_{-1}^1 d \cos{\theta} \int _0^{2\pi} d\varphi\, 
\left({\bf Y_l^{m}}(\theta,\varphi)\right)^{\star}
\, {\bf Y_{l^{\prime}}^{m^{\prime}}}(\theta,\varphi)\, 
= \delta_{l, l^{\prime}} \; \delta_{m, m^{\prime}} \; .
\eeq
When considering the interference of ${\cal A}^{\xi}(J=2)$ 
with the scalar exchange amplitude ${\cal A}(J=0)$, 
only the last component in ${\bf Y_0^0}(\theta,\varphi)$ of the 
graviton amplitude
[that is proportional to $(1-3\xi)$] survives 
after angular integration, for equal
initial and equal final helicities. 
Then, the coefficient of 
this residual component  vanishes only in the case of a  massive 
graviton propagator, for which $\xi=\frac{1}{3}$. 
In the Einstein theory ($\xi=\frac{1}{2}$), the coefficient
does not vanish, and it is responsible for the 
non-orthogonality of the graviton and scalar  
amplitudes.

By summing  
the graviton-scalar interference obtained starting from the amplitudes
in Eqs.(\ref{ampz}) and (\ref{amplitudes})
over the external particles helicities,
one easily recovers the results in Eqs.(\ref{feru}) and (\ref{ferd}) 
obtained by summing the interference over the external polarizations.

On the basis of Eqs.(\ref{ampu}) and (\ref{amplitudes}),
it is now straightforward to verify that
there are not problems with angular momentum selection rules,
as far as the interference of the graviton amplitudes and
the vector-boson ($J=1$) exchange amplitudes are concerned.
For the sake of  completeness, we present in the Appendix II
the corresponding results 
for $ \ci^m_{a,b}(2,1)$ and  $\Delta_{a,b}(2,1)$ ,
for all the external fermion and vector-boson states 
considered 
for the graviton-scalar interferences.

\section{Conclusions}
Selection rules for angular momentum conservation 
have been considered in the
framework of quantum gravity.
As required by angular momentum conservation,
the interferences of  $s$-channel amplitudes 
mediated by particles with different spins $J=0,1,2$ must
vanish after complete angular integration on the final state.
We find that, in the case of a propagating {\it massive} graviton,
these selection rules are satisfied for any  graviton mass.
On the contrary, as a consequence of the vDVZ discontinuity 
(for which the massless limit of massive gravity 
is different from the Einstein theory),
the interferences of $J=0$ and $J=2$ amplitudes do not vanish
in the {\it massless} gravity, whenever all the external states are massive. 
We checked this property in the $s$-channel $p_1p_2\to p_3p_4$ scatterings, 
where initial and final states  are either fermions or gauge bosons.
We conclude that angular momentum selection rules in 
the quantum gravity of the Einstein theory are broken.

This result could be interpreted in the following way.
Assuming angular momentum conservation at each interaction vertex,
 a massless graviton propagator behaves 
as if it was carrying a further scalar degree
of freedom coupled to the masses of matter fields with gravitational strength.
This extra scalar field would not decouple in  physical processes, 
leading to the breaking of angular momentum selection rules.

The latter interpretation would anyhow be   in contrast
with unitarity and the energy-momentum tensor conservation,
since, in the processes considered, only 
the spin-2 transverse polarizations (with helicities $\lambda =\pm 2$) are 
exchanged in the massless graviton propagator.

Then, we conclude that, in the Einstein 
theory, angular momentum is not conserved at quantum level 
in the graviton coupling to massive matter fields, even if the total
angular momentum is conserved in the scattering process.
This effect could be interpreted as a new kind of quantum anomaly.
In this regard, the {\it massive} quantum gravity, or even its massless limit,
is a better-behaved theory, being {\it anomaly free}.

The present results could be due to the
use of perturbation theory around the flat metric.
Then, the breaking of angular momentum selection rules 
could  simply suggest that the standard  approach to perturbation 
theory in quantum gravity is not completely consistent.

On the other hand, assuming that quantum gravity based on 
the Einstein theory correctly describes the gravitational
interactions, 
the present breaking of angular momentum selection rules
seems to be connected to a new quantum effect that should show up 
in some physical process. In particular, it
 could {\it in principle} be 
measured by some experiment (although unrealistically at the moment), 
if the Higgs boson will be discovered.

\section*{Acknowledgments}
We acknowledge useful discussions with M. Giovannini, C. Montonen, 
M. Porrati, M. Testa,
and G. Veneziano. A.D. and E.G. also thank Academy of Finland 
(project number 48787) for financial support.

\newpage
\section*{Appendix I}
\vskip 1cm
\begin{itemize}

\item {\bf \large Feynman Rules}

The Feynman rules used in this paper are the following
\cite{QG}
\bea
{\bf H-\bar{f}-f}\, &=&\, -i\, \lambda_f \; ,
\nonumber\\
{\bf H-W^+_{\alpha}-W^-_{\beta}}\, &=&\,i\, g_W m_W \, g_{\alpha\beta}  \; ,
\nonumber\\
{\bf V_{\mu}-\bar{f}-f}\, &=&\, 
\frac{i}{2}\left(g^f_V\, \gamma_{\mu}-g^f_A\, \gamma_{\mu}\gamma_5\right) \; ,
\nonumber\\
{\bf V_{\mu}(q)-W^+_{\alpha}(p^+)-W^-_{\beta}(p^-)}\, &=&\,i\, g_W
\left\{ 
g_{\mu\alpha}\left(q_{\beta}-p^+_{\beta}\right)+
g_{\mu\beta}\left(p^-_{\alpha}-q_{\alpha}\right)+
\right. \nonumber\\
&+& \left. g_{\alpha\beta}\left(p^+_{\mu}-p^-_{\mu}\right)\right\} \; ,
\nonumber\\
{\bf G_{\mu\nu}-\bar{f}(k_2)-f(k_1)}\, &=&\, 
-\frac{i}{4 M_P}\left\{
W^{(f)}_{\mu\nu}(k_1,k_2) +W^{(f)}_{\nu\mu}(k_1,k_2) \right\} \; ,
\nonumber\\
{\bf G_{\mu \nu}-W^+_{\alpha}(k_1)-W^-_{\beta}(k_2)}\, 
&=&\, 
-\frac{i}{M_P}\left\{
W^{(V)}_{\mu\nu\alpha\beta}(k_1,k_2)+W^{(V)}_{\nu\mu\alpha\beta}(k_1,k_2) \; ,
\right\}
\nonumber
\eea
where
\bea
W^{(f)}_{\mu\nu}(k_1,k_2)&=&
\gamma_{\mu}\left(k_{1\nu}+k_{2\nu}\right)-
\eta_{\mu\nu}\left(k_1 \!\!\!\!\!/ + k_2  \!\!\!\!\!/ -2 m_f\right)
\nonumber\\
W^{(V)}_{\mu\nu\alpha\beta}(k_1,k_2)&=&
\frac{1}{2}\eta_{\mu\nu}\left(k_{2\alpha}k_{1\beta} -
\eta_{\alpha\beta}\, k_1\cdot k_2\right)+
\eta_{\alpha\beta} k_{1\mu}k_{2\nu}-\eta_{\mu\beta} k_{1\nu}k_{2\alpha} \; ,
\nonumber \\
&+& \eta_{\mu\alpha}\left(\eta_{\nu\beta}\, k_1\cdot k_2
-k_{2\nu}k_{1\beta}\right)
+m_W^2\left(
\eta_{\mu\alpha}\eta_{\nu\beta}
-\frac{1}{2}\eta_{\mu\nu}\eta_{\alpha\beta}
\right) \; .
\nonumber
\eea
Above, $p \!\!\!/ = \gamma^{\alpha} p_{\alpha}$,
 $M_P$ is the reduced Planck mass, defined as 
$M_P^2=(8\pi G_N)^{-1}$ (where $G_N$ is the Newton constant), and
$m_f$, $m_W$ are the fermion, vector-boson masses,
respectively. $\; V_{\mu}$, $H$, and $G_{\mu \nu}$
are a neutral vector gauge boson, Higgs boson and graviton fields,
respectively.
The momenta  in the $G$-$W$-$W$  Feynman rule
are entering into the vertex, while in
$G$-$\bar{f}(k_2)$-$f(k_1)\;$, $f(k_1)$ / $\bar{f}(k_2)$ 
stands for an incoming/outgoing fermion $f$ of momenta 
$k_1$ / $k_2$, respectively.

The corresponding vertices for the $W^{\prime}$ vector boson, 
are obtained 
just changing $g_W\to g_{W^{\prime}}$ and $m_W\to m_{W^{\prime}}$.
\newpage
\item {\bf \large Spherical Harmonics}

The spherical harmonics ${\bf Y_l^m}(\theta,\varphi)$ are eigenstates 
of the angular momentum operator 
${\bf \hat{L}^2}$ and its projection on the $z$ axis ${\bf \hat{L}_z}$,
satisfying \\
${\bf \hat{L}^2} {\bf Y_l^m}=
l(l+1)\, {\bf Y_l^m}$ and
${\bf {\hat L}_z} {\bf Y_l^m}=m {\bf Y_l^m} \; $. 
Below, we report explicitly the  spherical harmonics
 entering into Eqs.(\ref{ampz})-(\ref{amplitudes})
\bea
{\bf Y_0^0}(\theta,\varphi)&=&\frac{1}{\sqrt{4\pi}} \; ,
\nonumber\\
{\bf Y_1^0}(\theta,\varphi)&=&\sqrt{\frac{3}{4\pi}}\cos{\theta} \; ,
\nonumber\\
{\bf Y_1^{\pm 1}}(\theta,\varphi)
&=&\pm \sqrt{\frac{3}{8\pi}}\sin{\theta}\, e^{\pm i\varphi} \; ,
\nonumber\\
{\bf Y_2^0}(\theta,\varphi)
&=&\sqrt{\frac{5}{16\pi}}\left(1-3\cos^2{\theta}\right) \; ,
\nonumber\\
{\bf Y_2^{\pm 1}}(\theta,\varphi)&=&\pm \sqrt{\frac{15}{8\pi}}\cos{\theta}
\sin{\theta}\, e^{\pm i\varphi} \; .
\nonumber
\eea
\end{itemize}
\section*{Appendix II }
In this appendix, 
we consider the interferences of the $J=2$ and $J=1$ amplitudes,
assuming the definitions in Eqs. (\ref{iim})-(\ref{disc}).
Terms arising from the 
axial-vector coupling of fermions are included, too, although
they do not give rise to any discontinuity.
\begin{itemize}
\item
${\bf e^+ e^- \to f\bar{f}}$
\bea
\ci^m_{e,f}(2,1)
&=& 2 g_V^eg_V^f\left\{
\beta_e\beta_f\left(r_f+r_e (1-\frac{4}{3} r_f)\right)\cos{\theta}
+\frac{1}{4}\beta_e^{3}\beta_f^{3}\cos^3{\theta}
\right\}
\nonumber \\
&-&\frac{g_A^e g_A^f}{4}\beta_e^2\beta^2_f\left(1-3\cos^2{\theta}\right) \; ,
\label{j1eeff}
\eea
and
\bea
\Delta_{e,f}(2,1)=-\frac{4}{3} g_V^eg_V^f \beta_e\beta_f r_e r_f \cos{\theta}\; .
\eea
In this case,
the discontinuity vanishes after total angular integration,
and is proportional 
to $r_e r_f \sim m^2_e m^2_f$, since it is connected to
 the traces of the energy-momentum 
tensors of the initial and final states.
In the limit of massless fermions, the interference does
not vanish. Indeed, contrary
to the $J=0$ channel, the $J=1$ channel has the same chirality structure 
as the $J=2$ channel, and  the $(J=1)-(J=2)$ interference
survives also in the massless fermion limit.

The orthogonality in  Eq. (\ref{j1eeff}) 
was first noticed in \cite{dgm1},  although the corresponding 
results were obtained
in a different context and in the massless fermion limit.
\item ${\bf e^+ e^- \to W^+W^-}$
\bea
\ci^m_{e,W}(2,1)\!\!
&=&\!\!-\frac{g_W g_V^e}{r_W}\left\{
\beta_e\beta_W\left(\frac{1}{4} -\frac{1}{3}r_e +
\frac{3}{2}r_W + 14r_er_W + 6r_W^2 - 8r_er_W^2\right)
\cos{\theta}\right.\nonumber\\
&+&\left. 
\beta_e^{3}\beta_W^{3}\left(-\frac{1}{4}+\frac{3}{2}r_W\right)
\cos^3{\theta}
\right\}
\eea
and
\beq
\Delta_{e,W}(2,1)=-\frac{g_W g_V^e}{3 r_W}
\beta_e\beta_W r_e \left(1-12 r_W^2\right)\, \cos{\theta} \; .
\eeq
In this case the contribution of the fermion 
axial coupling exactly vanishes. 
The $r_e/r_W$ dependence in the discontinuity arises from 
terms proportional 
to $1/m_W^4$ in the sum over polarizations of the two final $W$'s, 
combined with the  terms $r_er_W$  emerging 
from the vDVZ discontinuity.
\item ${\bf W^+ W^- \to W^{\prime +} W^{\prime -}}$
\bea
\ci^m_{W,W^{\prime}}(2,1)
&=& -\frac{g_W g_{W^{\prime}}}{r_Wr_{W^{\prime}}}
\beta_W\beta_{W^{\prime}}\left\{\left(
-\frac{1}{48}+\frac{7}{4}r_W-r_W^2
+\frac{45}{4}r_Wr_{W^{\prime}} + 42 r_W^2r_{W^{\prime}}
\right.\right.\nonumber \\
&-& \left.\left. 12r_W^2r_{W^{\prime}}^2
\right)\, \cos{\theta}
+
\beta_W^{2}\beta_{W^{\prime}}^{2}
\left(\frac{1}{16}+\frac{9}{4}r_W r_{W^{\prime}}
-\frac{3}{4}r_W\right)
\cos^3{\theta}\right\}
\nonumber \\
&+&\left( r_W\, \leftrightarrow\,  r_{W^{\prime}}\right)
\eea
and 
\beq
\Delta_{W,W^{\prime}}(2,1)=
\frac{g_W g_{W^{\prime}}}{12 r_Wr_{W^{\prime}}}
\beta_W\beta_{W^{\prime}}\left(1-12 r_W^2\right)
\left(1-12 r_{W^{\prime}}^2\right)\, \cos{\theta}
\eeq
\end{itemize}

In the above equations, 
$g^e_V$ and  $g^e_A$ are the vectorial and axial coupling of 
fermions to the neutral gauge boson $V$, and  $g_W$ and  $g_{W^{\prime}}$ 
are the  couplings of the gauge bosons 
$W^{\pm}$ and $W^{\prime \pm}$ to $V$, respectively
(cf. Appendix I).
\newpage

\end{document}